\documentclass[prd,showpacs,amsmath,showkeys,twocolumn,floatfix]{revtex4} 
\usepackage{epsfig,dcolumn}
\usepackage{graphicx}
\usepackage[usenames]{color}
\usepackage{graphicx}
\usepackage{bm}

\def\x{{\bf x}}
\def\y{{\bf y}}

\def\k{{\bf k}}
\def\q{{\bf q}}
\def\p{{\bf p}}

\def\A{{\bf A}}
\def\B{{\bf B}}

\def\lsim{\mathrel{\rlap{\lower4pt\hbox{\hskip1pt$\sim$}}
    \raise1pt\hbox{$<$}}}
\def\gsim{\mathrel{\rlap{\lower4pt\hbox{\hskip1pt$\sim$}}
    \raise1pt\hbox{$>$}}}
\begin{document}

%\preprint{IFT/12/02}

%\preprint{APS/123-QED}

\title{Chiral symmetry restoration and deconfinement in Coulomb Gauge QCD}
% Force line breaks with \\

\author{ Peng Guo and Adam P. Szczepaniak}
\affiliation{ Physics Department and Nuclear Theory Center \\
Indiana University, Bloomington, IN 47405, USA. }

\date{\today}% It is always \today, today,

\begin{abstract} 
 In the framework of Coulomb Gauge QCD we explore dynamical
  breaking of chiral symmetry and screening of the  confinement potential at finite density. The screened potential is applied in study of charmonium dissociation. 
  \end{abstract} 

\pacs{11.10.Ef, 12.38.Mh, 12.40.-y, 12.38.Lg}

\maketitle

\section{Introduction}
Since the recent discovery of the strongly correlated QCD fluid, the  phenomenology of  quark-gluon plasma (QGP) has attracted a lot of 
attention~\cite{Adams:2005dq}.  Many predictions have been made for the several possible phases of the QGP in which quarks and gluons are no longer confined to nucleons and pions~\cite{Rischke:2003mt,Gyulassy:2004zy,Harris:1996zx,Gross:1980br}.  At extremely high temperature and low density, thermal excitations of the gluon field are expected to screen interactions between color charges and ultimately, due to asymptotic freedom  result in a weakly interacting quark gas~\cite{Collins:1974ky,karsch,Laermann:2003cv}. Similarly, at low temperature but high density Debye screening is expected to reduce the range of strong interactions. Novel phases that correlate quark color and flavor are predicted to occur at asymptotic densities due to the attractive nature of quark-quark interactions in certain color-flavor locked combinations~\cite{rajagopal,wilczek}.  
 Thus  finite  temperature and/or density are expected to reflect on various aspects  of confinement. 
  Furthermore any  modification of the ground state influences symmetry  properties and in particular restoration of 
    chiral symmetry is expected. At finite density and low temperature  the precise relation between chiral symmetry restoration and deconfinement is not yet known. A common wisdom is that in this regime there is a phase transition from hadronic (confined) matter at low density to the unconfined (possibly superconducting) phase at higher densities~\cite{Davis:chiral, Galina,Kocic,Glozman:2007tv}. Recently, however it has been observed that the confined and deconfined phases may  be separated by a phase where quarks are confined but  chiral symmetry is restored. 
  In the limit of an infinite number of colors $N_C \to \infty$ this so called quarkyonic phase would in fact extend to infinite density, since in this limit Debye screening due to quark loop vanishes~\cite{mclerran}.   In this paper we examine the possible emergence of this new phase using a canonical formulation of the QCD many body problem in the Coulomb gauge.  The Coulomb gauge  canonical  formulation can describe both  finite temperature and density. In this formulation  manifestation of deconfinement  can be inferred from the temperature and/or density dependence of the color Coulomb interaction.  In contrast lattice simulations at finite density are still at their  infancy~\cite{Fodor:2001au,Fodor:2001pe}. 
One of the smoking gun signals of deconfinement is the possible  dissociation  of heavy 
quarkonia~\cite{Doring:lat,Kaczmarek:lat,Petreczky:lat,satz,wong,digal,Mocsy:spec2007,Mocsy:spec2008}.  By studying density dependence of the Coulomb gauge heavy quark potential  we will be able to explore charmonium properties at  finite density. 

The paper is organized at follows. In the following section we discuss the Coulomb gauge QCD and approximations relevant to the problem in hand.   Zero and finite density properties are discussed in Sections~\ref{q0} and ~\ref{qf}, respectively.  Charmonium dissociation  is studied in Sec.~\ref{jp} and followed by conclusions and outlook. 

\section{ Coulomb gauge QCD} 
In this section we briefly discuss QCD in the Coulomb gauge and the approximations appropriate for the high density and/or temperature 
systems~\cite{Szczepaniak:2001rg,Szczepaniak:2003ve}. In the Coulomb gauge gluons are described by the transverse potentials, $\A^a(\x), a = 1\cdots N_C^2 -1$, $\bm{\nabla} \cdot \A^a(\x) = 0$   and the conjugated, transverse momenta, $\bm{\Pi}^a(\x)$, 
\begin{equation}
[ \A^a(x), \bm{\Pi}^b(\y) ] = i\bm{\delta}_T(\x - \y)\delta^{ab},
\end{equation}
 where  $\bm{\delta}_T(\x-\y) \equiv [{\bf I}  - \bm{\nabla} \bm{\nabla}/\bm{\nabla}^2] \delta^3(\x-\y)$. 
 The canonical momentum $\bm{\Pi}^a(\x)$ is the  negative of the transverse component of the chromo-electric field. 
 The quark and antiquark  degrees of freedom will be defined below in terms of the canonical set of Dirac fields, $\psi_i(\x),\psi^{\dag}_i(\x)$, $i=1\cdots N_C$, one for each flavor. The Hamiltonian is given by 
 \begin{equation} 
 H = H_D + H_{YM} + H_C, \label{H}
 \end{equation} 
 where $H_D$ contains the  Dirac kinetic energy and quark-transverse gluon interaction, $H_{YM}$ is the Yang-Mills term, which contains the gluon kinetic energy and  the three- and four-gluon interactions and  finally $H_C$ is the Coulomb potential given by,
 \begin{equation} 
H_C =  {1\over 2} \int d\x d\y  {\cal J}^{-1} \rho^a(\x) K(\x,a;\y,b)[\A] {\cal J} \rho^b(\y). 
\end{equation} 
 It represents the non-abelian Coulomb gauge interaction between color charge densities, $\rho^a(\x)  = \psi^{\dag}(\x) T^a \psi(\x) 
   + f_{abc}  \bf{A}^b(\x) \bm{\Pi}^c(\x) $, mediated by the  Coulomb kernel $K[\A]$  given by 
\begin{equation}
K(\x,a;\y,b)[\A] = \left[ {g \over {\bm{\nabla} \cdot {\cal D} }} (-\bm{\nabla}^2) {g \over { \bm{\nabla} \cdot {\cal D}}}  \right] _{(\x,a;\y,b)}. \label{Keff}
\end{equation} 
Finally ${\cal J} = \mbox{Det}\bm\left[-{\nabla}\cdot  {\cal D}\right]$ is the determinant of the Faddeev-Popov operator; ${\cal D} = {\cal D}_{ab} = \delta_{ab} \bm{\nabla}  + g f_{acb} {\bf A}^c$  is the covariant  derivative in the adjoint representation, and $\B$ is the chromo-magnetic field, $\B^a(\x) =  \bm{\nabla} \times \A^a(\x) + (g/2) f_{abc} \A^b(\x) \times \A^c(\x)$. At  every  space point-$\x$ and color component $a$ the Coulomb gauge potentials, $\A^a(\x)$ are an analog of a curvilinear coordinate. This is because their values are restricted to reside within the boundary of the Gribov region which has a nontrivial metric determined by ${\cal J}$~\cite{Christ:1980ku}. A confinement scenario in the Coulomb gauge 
  states that it is the field configurations near the boundary of the Gribov region $\partial \Omega$  that dominate the QCD 
   vacuum~\cite{Zwanziger:1993dh,Gribov:1977wm}  and it follows that   fluctuations near the boundary lead to massive quasiparticle 
   excitations.  Since the exact parametrization of the Gribov horizon is not known the quantitative description of this confinement scenario varies depending on how restriction to the Gribov horizon is implemented, but the general features seem to be robust~\cite{Epple:2007ut,Epple:2006hv,Reinhardt:2004mm,Feuchter:2004mk}. With this picture in mind we approximate the ground state of a finite density quark plasma by the state with no quasiparticle gluon excitations and the large background fields concentrated near $\partial \Omega$ lead to enhancement in the long range  behavior of the Coulomb 
   kernel~\cite{Szczepaniak:2001rg,Feuchter:2004mk}. That is we make the replacement 
   \begin{equation} 
   K(\x,a;\y,b)[\A]  \to \langle K(\x,a;\y,b)[\A]  \rangle = K(\x-\y)\delta_{ab} 
   \end{equation} 
with the potential $K(r)$ modified from its free ($\A=0$) form $K(r) = \alpha/r$, for large $r$ due to the large fields $\A \in \partial \Omega$. In particular  
we approximate the kernel  by the from 
\begin{equation}
K(r) = K_C(r) + K_L(r) 
\end{equation} 
with $K_C$ and $K_L$  being the short-range Coulomb and long-range linear potentials, respectively which will be discussed in detail in the following section.  The final Hamiltonian, describing massless quarks with energies  below gluon quasiparticle  excitations  is given by 
\begin{eqnarray}
H  &=& \int d\x \psi^{\dag}(x) (-i \bm{\alpha} \cdot \bm{\nabla}
) \psi(x) \nonumber \\
& + & \frac{1}{2} \int d\x d\y \psi^{\dag}(x) T^a \psi(x) K(|\x - \y|) \psi^{\dag}(y) T^a \psi(y).  \nonumber \\ \label{Heff} 
\end{eqnarray}
Note, that from Eq.~(\ref{Heff}) it follows  that it is $-C_F K(r)$ which is the instantaneous interaction in the color-singlet $q\bar q$ channel. 

\section{Quarks at zero density} 
\label{q0} 
\begin{figure}[h]
\begin{center}
\includegraphics[width=2.4 in,angle=270]{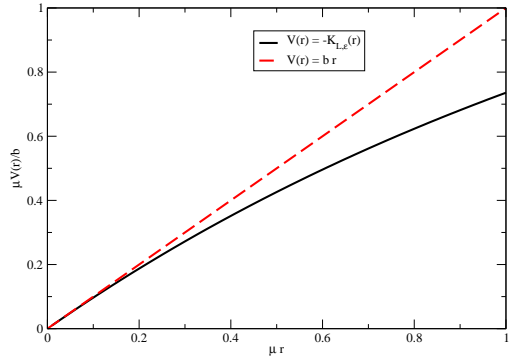}  
\caption{ IR finite approximation to the linear potential of Eq.~\ref{k0}. The proper potential for describing color-nonsinglet state, ({\it cf.} Eq.~(\ref{k0})) corresponds to a  downward-shift by two units. \label{kl}} 
\end{center} 
 \end{figure}

In presence of the effective  density-density interaction mediated by the kernel  $K$, in Eq.~(\ref{Heff}) quarks and antiquarks acquire effective mass, which in the mean filed approximation can be described within the Hartree-Fock-Bogolubov framework~\cite{Davis:chiral, Galina,Kocic, Yaouanc}.  The single quark quasiparticle operators are defined by a canonical transformation to a plane wave representation of the Dirac fields. 
\begin{eqnarray} 
\psi(\x)  &=&\sum_\lambda \int d\k 
 e^{i \k\cdot\x }  [u(\k,\lambda) b_{\k,\lambda}  +
v(-\k,\lambda) d^{ \dag }_{\k,\lambda}]  \nonumber \\
\end{eqnarray}
and similarly for $\psi^\dag$. Here $d\k \equiv d^3k/(2\pi)^3$,  $\lambda=\pm1/2$ is the quark (antiquark) spin projection and $b (d), b^{\dag}(d^{\dag}) $ are the quark (antiquark) annihilation  and creation operators, respectively.  These  quasiparticle  operators satisfy the standard fermion  anti-commutation  relations and define the vacuum state, by $b_{\lambda,\k}  |\mbox{vac} \rangle = d_{\lambda,\k}  |\mbox{vac} \rangle=0$. The single particle wave functions are given by 
\begin{eqnarray} 
&& u^T(\lambda,\k) =  \frac{1}{\sqrt{2}} \left(  \sqrt{1+ \frac{E_k}{m_k} }   \chi_{\beta}, 
\sqrt{1- \frac{E_k}{m_k} }\sigma \cdot \hat{\k}  \chi_{\beta} \right) \nonumber \\
&& v^T(\lambda,\k)= \frac{1}{\sqrt{2}}    \left( 
  \sqrt{1- \frac{E_k}{m_k} }   \sigma \cdot\hat{\k} \bar{\chi}_{\beta} , 
  \sqrt{1+\frac{E_k}{m_k} }    \bar{\chi}_{\beta} \right),  \nonumber \\
\end{eqnarray}
where $\bar{\chi} \equiv i \sigma_{2}  \chi $,  
  $m_k/E_k \equiv  \sin\phi_k$  and  $ \phi_k$ is the  BCS  angle, which determines the number density of quark-antiquark  pairs in the BCS vacuum, {\it i.e.} the quark condensate and the extent of chiral symmetry breaking~\cite{Adler:1984ri,Le Yaouanc:1983iy,Bicudo:1989sh}. At zero temperature and density the BCS angle is determined by minimizing the vacuum energy density, $\delta \langle\mbox{vac} | H | \mbox{vac} \rangle /\delta \phi_{k}=0$, which leads to the gap equation ($s_p \equiv \sin\phi_p, c_p \equiv \cos\phi_p$)
\begin{equation} \label{gap}
  p s_p   = 
  \frac{C_F }{2} \int d\k \tilde  K(|\k-\p|)  [ s_k c_p - c_k s_p\hat{\k} \cdot \hat{\p} ].  
\end{equation}
 Here $\tilde K(p)$ is the Fourier transform of the effective potential from Eq.~(\ref{Keff}). The Fourier transform of the linear potential, $K_L$ has to be taken with care, since  naively, $\int d\x |x| \exp(-\k\cdot \x)  = \infty$. We introduce an infrared regulator, $\epsilon$ and define~\cite{Yaouanc:1984}
\begin{equation} 
K_L(r) \to K_{L,\epsilon}(r) = \frac{2b}{\epsilon^2} \left(\frac{1}{R} - \frac{e^{-\epsilon R}}{R} \right)  - \frac{2b}{\epsilon}  \label{k0} 
\end{equation} 
so that $\lim_{\epsilon \to 0} K_{L,\epsilon}(r) = -br$. The difference between the liner potential and the IR finite approximation is shown in Fig.~\ref{kl}. In momentum space the IR finite  kernel becomes 
\begin{equation} 
\tilde K_{L,\epsilon}(p) = \frac{8 \pi b}{p^2(p^2 + \epsilon^2)} - \frac{2b}{\epsilon} (2 \pi)^{3} \delta^{3}(\p)  \label{k1} 
\end{equation} 
It is clear that the $\delta$ term does not contribute the gap equation~(\ref{gap}) and the gap equation is well defined in the limit $\epsilon\to0$. Equivalently, the gap equation is invariant under a constant shift in the potential,
\begin{equation} 
K(r) \to K(r) + C. \label{shift} 
\end{equation} 
Since such a shift induces a contribution to the Hamiltonian  proportional to the square of the total charge  operator $\sum_a Q^aQ^a$, 
\begin{equation} 
Q^a = \int d\x \rho^a(\x) 
\end{equation} 
invariance under~(\ref{shift}) is an exclusive property of color singlet states and it is only matrix element invariant under global color rotations 
 that are  are physical. In contrast, under this shift,  the  energy $\Omega_p$  of a single quark state ( $b^{\dag}(\lambda,\p) |\mbox{vac} \rangle$) that is given by, 
\begin{equation} 
\Omega_  p  =  p c_p +   \frac{C_F }{2} \int d\k  \tilde K(|\k-\p|)  [ s_k s_p + c_k c_p\hat{\k} \cdot \hat{\p} ] \label{omega}
\end{equation}
transforms to 
\begin{equation} 
\Omega_p \to \Omega_p + \frac{C_F}{2} C . 
\end{equation} 
So single quark states are clearly unphysical. For $\epsilon \to 0$ the $\delta$-term in $\tilde K$  dominates the integrand in Eq.~(\ref{omega}) and the quark self-energy  becomes negative and tends to $-\infty$ in the $\epsilon \to 0$ limit. As pointed out in \cite{Yaouanc:1984} this is necessary in order for color-singlet $q\bar q$ excitations  to have finite, non-negative energies. This is because the potential energy in the $q\bar q$ bound state, given by $-C_F \tilde K(p)$, where $p$ is the relative momentum between the quark and the antiquark  is large  and positive for small $\epsilon$ (and approached $+\infty$ in the limit $\epsilon \to 0$). Thus the infinities in the $\epsilon \to 0$ limit  cancel between the self-energies and the residual interaction between the quark and the  antiquark. Since $\phi\ne 0$ is a lower energy state compared to $\phi=0$, after cancelation of the IR divergencies, the  finite energy of color single excitations is $C$-independent and non-negative. This is, however not the case for color-nonsinglet states. For example a single quark state with energy given by Eq.~(\ref{omega}) has negative energy which becomes $-\infty$ in the limit when the interaction is confining ({\it i.e.} $\epsilon \to 0$).  This is clearly unphysical as one would expect colored states to have positive, and IR diverging energies, $\Omega \to + \infty$ in the confining limit. Since the  shift in Eq.~(\ref{shift}) is a symmetry of the physical sector we can choose $C$ to cancel the IR divergence in the quark self energy, and redefine $K_L$ accordingly 
  Thus instead of Eq.~(\ref{k1}) we should  use 
\begin{equation} 
\tilde K_{L,\epsilon}(p) =  \frac{8 \pi b}{p^2(p^2 + \epsilon^2)}, \label{k01} 
\end{equation} 
which is positive, gives $\Omega \to + \infty$ as $\epsilon \to 0$ and does not affect properties of color singlet states. 
In fact  and interaction  without the IR divergent constant term constant term is obtained  when computing the expectation value of $K[A]$ ({\it cf.} Eq.~(\ref{Keff})) in a mean filed ansatz  for the gluon vacuum distribution~\cite{Szczepaniak:2001rg,Szczepaniak:2003ve}.  

A similar argument for regulating the $\p=0$ singularity of $\tilde K_L$ was proposed in ~\cite{LeYaouanc:1987ct}. There by explicitly restricting the  spectrum of the Hamiltonian to include only the  color-singlet subspace   an even stronger  constraint on the  IR momentum dependence of the kernel was derived, namely $\tilde K_L(0) = 0$.  With an IR regulated (finite-$\epsilon$) kernel the minimal  physical requirement, however is that colored  excitations have positive energies, (becoming $+\infty$  as $\epsilon \to 0$). Thus  we do consider kernels $\tilde K(p)$ which have integrable, but in principle finite zero-modes. 

\section{Quarks at finite density}
\label{qf} 
With the interactions in the Hamiltonian now well defined in the IR,  the system at finite quark density can be described using standard many body techniques~\cite{Davis:chiral,Galina, Kocic}. In particular the finite density gap equation becomes, 
\begin{equation} 
\label{gapf}
 p s_p  = \frac{C_F }{2} \int d\k F_k  \tilde{K}_{eff}(|\k - \p |)
  [ s_k c_p - c_k s_p \hat{\k} \cdot \hat{\p} ] 
\end{equation}
and the single quark energy is given by, 
\begin{equation} \label{selfenergyf}
  \Omega_p  = p c_p +  \frac{C_F}{2}  \int d\k  F_k 
 \tilde{K}_{eff}(|\k -\p |)  [ s_k s_p   +  c_k c_p  \hat{\k} \cdot \hat{\p}].  
\end{equation}
Here $F_k = 1 - n_k  - \bar{n}_k$ and $n_k(\bar{n}_k)$ is  the quark (antiquark) occupation number at zero-temperature,
\begin{equation} 
n_k = \theta( \Omega_k - k_F), \; \bar{n}_k = \theta(\Omega_k + k_F) 
\end{equation} 
with $k_F$ denoting the the Fermi momentum. Because of the Pauli blocking factor $F_k$ interactions between quarks have small effect on quark levels inside the  Fermi  sphere where quarks are effectively free $\Omega_p \sim pc_p $. The self-energy contributes mainly for states above the Fermi surface and results in states which are  similar to the confined quark states at zero-density. Since any modification of the antiquark distribution from its vacuum value is suppressed at  finite quark density,  it will be ignored in the following and we will only consider quark-hole excitations near the quark Fermi surface.

\subsection{Screening effect at finite density }
\begin{figure}[h]
\begin{center}
\includegraphics[width=1.5 in,angle=0]{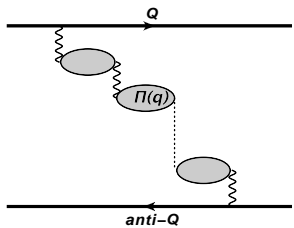}
\caption{  Vacuum polarization $\Pi_{q}$. \label{vac:bubble}} 
\end{center} 
 \end{figure}

At finite density the quark-quark potential $C_F \tilde{K}(p)$  is screened by particle-hole excitations near the Fermi surface, resulting in the effective interaction   $C_F \tilde{K}_{eff}(p)$  which enters Eqs.~(\ref{gapf}),(\ref{selfenergyf}),    
\begin{eqnarray}
\tilde{K}_{eff}(p) &= &   \tilde{K}(q) -
\tilde{K}(q)  \Pi (q) \tilde{K}(q)+ \cdots \nonumber \\
& = &   \tilde{K}(q) -\tilde{K}(q) 
\Pi(q)\tilde{K}_{eff}(q), 
\end{eqnarray}
or 
\begin{equation} \label{veff}
\tilde{K}_{eff}(q) = \frac{ \tilde{K}(q)}{1 + \tilde{K}(q) \Pi(q)}.
\end{equation}
The vacuum polarization, $\Pi(q)$ is shown in Fig.~\ref{vac:bubble} and it describes the probability for creating a particle-hole excitation. Since the  pair is excited in the colored state in the confining limit ($\epsilon \to 0$)  it is expected that $\Pi(q) \to 0$, since the amplitude is proportional the the inverse of the particle-hole excitation energy, which becomes infinite in the confining limit. Assuming phase transition does take place, however we can examine the effect of Debye screening in the  deconfined  phase near or above the phase transition density.  In this case vacuum polarization is given by 
\begin{eqnarray}
\Pi(q) &= &  -  \frac{n_{f}}{2} 
 \int  d\k   \frac{ n_k- n_{|\k+\q|}}{\Omega_k-\Omega_{|\k+\q|}}   \label{vacpol}   \\
& & \times \left[   1+ s_k s_{|\k+\q|} + c_k c_{|\k+\q|} \frac{\k\cdot(\k + \q)}{k(|\k+\q|)}  \right]  \nonumber 
\end{eqnarray}
where $n_f$ is the number of light flavors. The set of equations, ~(\ref{gapf}),~(\ref{selfenergyf}), ~(\ref{veff}) and ~(\ref{vacpol}) with $\tilde K(p)$ given by 
\begin{equation} 
\tilde{K}(p) = \frac{ 4\pi \alpha(p)}{p^2} + \frac{8 \pi b}{p^2(p^2 + \epsilon^2)}  \label{kfinal} 
\end{equation} 
forms a set of coupled equations which we solve numerically and discuss below. 

\subsection{Numerical result}
\label{numerical} 
\begin{figure}[h] 
\begin{center}
\includegraphics[width=2.8 in,angle=270]{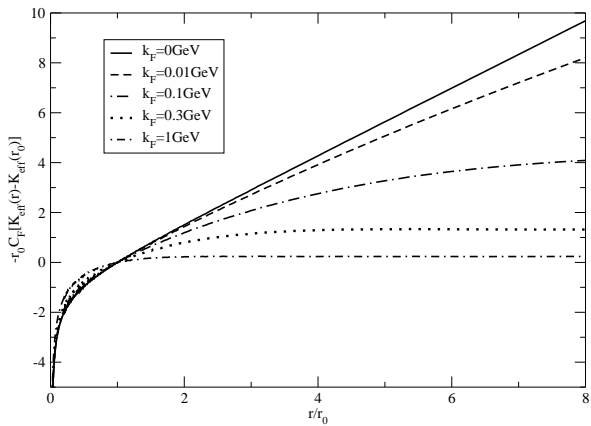} 
\caption{ Effective potential at finite density ($r_{0}= 1/0.45 GeV^{-1}$).  \label{fig:V}} 
\end{center} 
 \end{figure}

\begin{figure}[h]
\begin{center}
\includegraphics[width=2.8 in,angle=270]{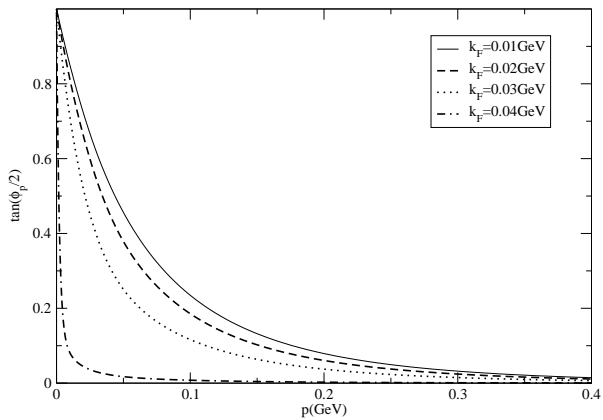}
\caption{ Solution of gap equation $\phi_{k}$  at finite density.  For $k_F \gsim 0.05 \mbox{ GeV}$ the solution of the gap equation is $\phi(p) =0$.  \label{fig:gap}} 
\end{center} 
 \end{figure}

\begin{figure}[h]
\begin{center}
\includegraphics[width=2.8 in,angle=270]{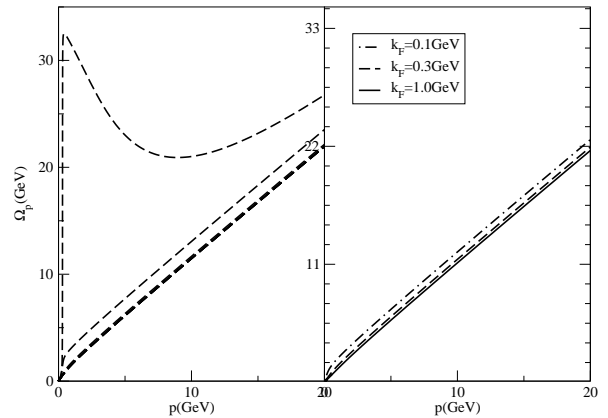}
\caption{ Left panel : Variation of single quark energy  $\Omega_{p}$ through iterations, for $k_F = 0.3\mbox{ GeV}$. 
$\Omega_{p}$ is the largest at first iteration. Right panel: $\Omega_{p}$ at the end of the iterations.  \label{fig:omega}} 
\end{center} 
 \end{figure}

\begin{figure}[h]
\begin{center}
\includegraphics[width=2.8 in,angle=270]{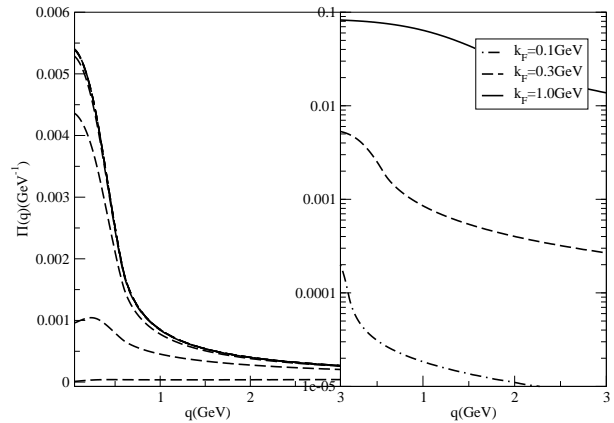}
\caption{ Same as in Fig.~\ref{fig:omega} for vacuum polarization $\Pi_{q}$. In  the left panel, $\Pi(q)$ is the smallest at first iteration. 
  \label{fig:pi}} 
\end{center} 
 \end{figure}

We solved coupled equations (\ref{gapf}), (\ref{selfenergyf}), (\ref{veff}) and (\ref{vacpol}) at finite density and zero temperature with two light quarks flavors 
using 
\begin{eqnarray}
\alpha(p) = \frac{4 \pi Z}{\beta^{\frac{3}{2}} \log^{\frac{3}{2}} (\frac{p^{2}}{\Lambda^{2}_{QCD}}+c) },
\end{eqnarray}
 with  $Z=5.94, c=40.68$ and $\Lambda_{QCD}= 250\mbox{  MeV}$  determined from fitting the zero-density $q{\bar q}$ potential~\cite{Guo:2007sm}. 
At zero density what makes quark energies finite is the  IR regulator $\epsilon$ in Eq.~(\ref{kfinal}), at finite density, above the deconfinement phase transition we expect, however, that the self-consistent set of equations will admit nontrivial  solutions in the limit $\epsilon \to 0$, with the Fermi momentum $k_F$ 
 taking over the role of the IR regulator instead.  We have verified numerically that this indeed is the case. The equations are solved by iterations. For given density, $k_F$ we start with a small, but finite $\epsilon$, {\it e.g.}   $\epsilon =0.008\mbox{ GeV}$, so the self-energy at the first iteration  has a sharp jump at the Fermi surface (see left panel in Fig.~\ref{fig:omega}). At this initial state 
 the vacuum polarization is highly suppressed (see left panel in Fig.~\ref{fig:pi}), particularly for  small momenta. After a few iterations, however,  the self-energy becomes  regulated  by the  
  vacuum polarization itself and simultaneously the  vacuum polarization increases at small momenta.  Finally, we reduce the initial value of the IR  
  regulator, $\epsilon$ and we verify that after several iterations solutions converge to the same value, regardless of the starting value of  the  regulator. 
   We repeat the calculations for several  values of the Fermi energy.  We show these finial results for the effective potential, BCS gap angle, quark single  energy and vacuum polarization
    in Figs.~\ref{fig:V},~\ref{fig:gap},~\ref{fig:omega},~\ref{fig:pi}, respectively. 
  
%  \color{red} Peng. Fig. showing the polarization tensor is missing. Also please split the fig showing $\Omega$ into a figure that shows how $\Omega$ changes throughout iteration (one figure) and the other that has the final plot of $\Omega$ for the same $k_F$'s at the other figures. \color{black} 

  %\color{red} Peng: Please redo the figures using xmgrace (and keep the source files) you should plot fewer values of $k_f$. \color{black} 
  
It is often stated that chiral symmetry restoration and deconfinement occur simultaneously  \cite{Davis:chiral, Galina}.
Our calculation illustrates that this need   not be the case. At finite density,  effective potential is already deconfined, but the gap equation admits nontrivial solutions.   Only when the Fermi momentum increases above, approximately  $k_{F} \sim 0.05$ to $0.06\mbox{ GeV}$,  the effective potential at large distance is not strong enough to sustain spontaneous chiral symmetry breaking. 
 
 \subsection{ Quarkyonic matter } 
 
 Another manifestation of the independence of chiral symmetry restoration and deconfinement is the appearance of confined,  chiraly  symmetric, quakyonic, matter~\cite{mclerran}. Since $K \propto g^2 \propto N_C$ we can extract explicit $N_C$ dependence of the effective interaction, 
    \begin{equation}\label{veffNc}
  \tilde{K}_{eff}^{N_C} = \frac{ \frac{3}{N_C}\tilde{K}^{N_C=3}}{1+ \frac{3}{N_C} \tilde{K}^{N_C=3} \Pi} .  
\end{equation}
The $q\bar q$ potential being proportional to $C_F K_{eff}^{N_C}$ at large $N_C$ becomes 
\begin{equation} 
C_F K_{eff} \to  \frac{3}{2} \tilde{K}^{N_C=3} 
\end{equation} 
since the vacuum polarization contribution is suppressed in the $N_C \to \infty$ limit. In this limit we thus find that the  Debye screening disappears and confinement is restored at any density. The gap equation (\ref{gapf}), however is not affected by the large $N_C$ limit, and the Pauli blocking remains in 
 effect. Thus  as density increases it will eventually prevent gap equation from developing nontrivial, charily broken solution~\cite{Glozman:2007tv}. 
%  This tell us chiral symmetry restoration doesn't have be associated to deconfinement. 

 \section{Charmonium binding at finite density}
 \label{jp}
\begin{figure}[h]
\begin{center}
\includegraphics[width=2.8 in,angle=270]{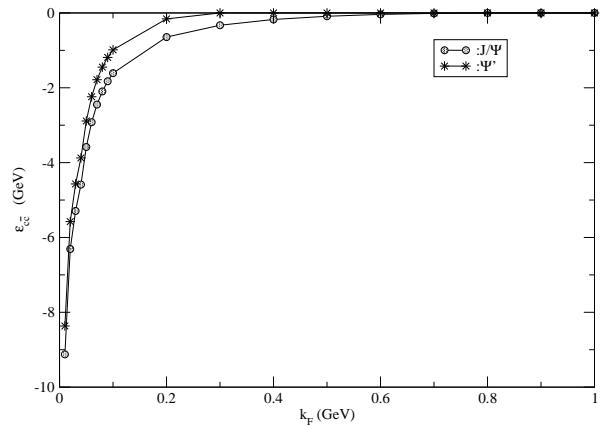}  
\caption{   $\epsilon_{J/\Psi}$ and $\epsilon_{\Psi'}$ at finite density, where $\epsilon_{J^{PC}} $ is defined in Eq.(\ref{schrodinger}).  \label{psi}} 
\end{center} 
 \end{figure}

\begin{figure}[h]
\begin{center}
\includegraphics[width=2.8 in,angle=270]{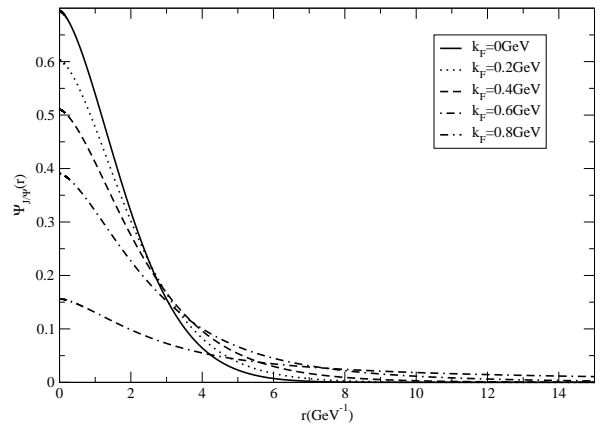}
\caption{   $J/\Psi$'s Wave function $\Psi_{J/\Psi}(r)$  at finite density.  \label{wf} } 
\end{center} 
 \end{figure}

Modifications of  charmonium  properties, in particular binding energy and  size may be  strongly affected when the bound state propagates 
  through the plasma.  At finite temperature  such modifications have been studied by computing charmonium spectrum using a temperature-dependent static potential    that is extracted from the lattice calculation~\cite{Doring:lat,Kaczmarek:lat,Petreczky:lat}. With such temperature dependent static potential, melting of charmonium  can be inferred from the temperature dependence of the  spectral function \cite{Mocsy:spec2008, Mocsy:spec2007} or by directly solving the bound state  Shr\"odinger equation~\cite{wong,bicudo}. In the former approach, melting of charmonium can be seen when the bound state peak of spectral function collapses and as temperature increases it becomes  buried under the continuous background.  From Shr\"odinger equation calculation melting  
   is inferred  at a temperature  when the  bound state solution  disappears. Melting temperature of the  $J/\Psi$,  from these two different approaches appears  to be in a very good  agreement at approximately  $1.6T_{c}$. 
   %\color{red} ( $T_C  $ is critical phase transition temperature between confinement and deconfinement phase, its value depends on the masses and the 
   %number of quark flavours. For pure SU(3) gauge theory without dynamical quarks, it is about  269 MeV. However, dynamical quarks will reduce this 
   %number dramatically, for two light as well as for two light plus one heavier quark flavor, it is about $175 \pm 10 MeV$. ref see \cite{satz} and F. Karsch and 
   %E. Laermann: hep-lat/0305025. So, for zero density, it is about 269 MeV in \cite{wong} and F. Karsch and E. Laermann.) \color{black} . 
   Here we use the Shr\"odinger equation 
   approach  to calculate dissociation of the $J/\Psi$ at finite density with  two flavors  of light quarks.  The mass $M_{J^{PC}}$ 
     of  the $J/\Psi$ is given by 
     \begin{eqnarray}
   M_{J^{PC}}  =  \epsilon_{J^{PC}}  + 2m + C_F[K_{eff}(r_{0}) -K_{CL}(r_{0})],
     \end{eqnarray}
    and
 \begin{eqnarray}   \label{schrodinger}
&&  ( \epsilon_{J^{PC}} - \frac{q^2}{m})\Psi_\alpha^N(q)   \\
&=& -C_F \sum_{\alpha'} \int d \mathbf{ q'} P_{L'_{q}}(\mathbf{ \hat{q}} \cdot \mathbf{ \hat{q}'}) \widetilde{K}_{eff}(|\mathbf{ q}-\mathbf{ q'}|) \Psi_{\alpha'}^N(q') .  \nonumber 
\end{eqnarray} 
with $N$ being the radial quantum number, $\alpha=(S_{q},L_{q})$ stands for total quark spin $S_q$ and relative orbital angular momentum $L_q$. The 
 heavy (charm) quark mass, $m$ absorbs any finite shift that has been removed from $K_{eff}$ as discussed in Sec.~\ref{q0}  and is fixed, 
  by fitting spin-averaged  $L_q=0$ charmonium  masses   $\bar{M}^{S}_{c \bar{c}}=\frac{1}{4}[M_{ 0^{-+}  }+3 M_{1^{--} }]  =3.068\mbox{ GeV}$. 
   %\begin{eqnarray}
%V_{eff}(r)= (4 \pi) \int_{0}^{\infty} \frac{p^{2} dp }{(2 \pi)^{3}} j_{0}(rp) \widetilde{V}_{eff}(p).
%\end{eqnarray}
The energy of $J/\Psi$ and its first radial excitation as a function of the light quark density are given in Table \ref{eigenenergy}  and the binding energy is plotted in
   Fig.~\ref{psi}. From our numerical evaluation is  follows that melting occurs at light quark density $k_{F} \sim 0.9 \mbox{ GeV}$ for $J/\Psi$ 
    and $k_{F} \sim 0.3 \mbox{ GeV}$ for $\Psi'$.  Finally in Fig.~\ref{wf}, we show how the wave function of $J/\Psi$ collapses with increasing density. 
   % and in Table.\ref{eigenenergy}, the average radius of $J/\Psi$ state is indicated too.
 
 \begin{table}[htdp]
\caption{ $J/\Psi$ and $\Psi'$  energy spectrum at finite density}
\begin{center}\label{eigenenergy}
\begin{tabular}{|c|c|c|c|} \hline
$k_{F}$(GeV)&  $J/\Psi$(GeV)& $\Psi'$(GeV)& $<r>_{J/\Psi} (GeV^{-1})$  \\  \hline
0.0 & 3.065 & 3.823 & $1/0.476$ \\ \hline
0.2  & 2.801& 3.288 & $1/0.417 $\\ \hline
0.4 & 2.771 & 2.946 & 1/0.351 \\ \hline
0.6 & 2.733 & 2.771  & 1/0.250 \\ \hline
0.8& 2.682 & 2.683 &  1/0.050  \\ \hline
\end{tabular} 
\end{center}
\label{default}
\end{table}

\section{Conclusion and outlook}
\label{sum}
 In this work  we studied the screened quark-quark  effective potential at finite temperature and presented numerical result for the effective potential. We investigated the restoration of chiral symmetry using the many body framework of the Coulomb gauge QCD and find  that the transition densities of chiral symmetry restoration and de-confinement need not be related.  In the deconfined phase, chiral symmetry can be broken at density $k_{F} \sim 0.05$ to $0.06 \mbox{ GeV}$. We also computed the melting density for $J/\Psi$ and $\Psi'$,  which we find to be at $k_{F} \sim 0.9 \mbox{ GeV}$ and 
  $k_{F} \sim 0.3 \mbox{ GeV}$ respectively.  Similar to the situation at high temperature and low density \cite{satz,wong, peskin,bhanot}, collision of gluons and quarks with heavy quarkonium will reduce the dissociation temperature of quarkonium. We are expecting the same situation occurs at high density and low temperature. As pointed out  previously  \cite{satz,wong, peskin,bhanot}, production of quarkonium in heavy-ion collision may provide clear signal for quark-gluon plasma. Clearly, the collision effect of gluons and quarks with heavy quarkonium should be incorporated in this case, the detail discussion of collision dissociation of heavy quarkonium at the phase of high density and low temperature is under way. 
  %
%  Recently  quark recombination scenario in Au+Au collisions has been discussed in  \cite{ayala} and it was  found that the transition probability   from free 
%quarks matter phase to a baryon gas phase is quite different from  that from free quarks matter phase to a meson gas phase as the function of energy 
%density. The former has a sharp jump whereas the latter transits smoothly with increasing density. 
%  The significant difference of  production rate between baryons and mesons from quark matter can be seen and measured in the proton to pion production 
%ratio as  a function of transverse momentum which has been shown in \cite{ayala}.  This scenario might be another approach to understand the quarks 
%matter phase and also to test phenomenon model computation.

\end{document}